\RequirePackage{fix-cm}
\documentclass[twocolumn,epjc3]{svjour3}
\smartqed  
\RequirePackage{graphicx}
\RequirePackage{mathptmx}      

\usepackage{cite}
\usepackage{appendix}
\usepackage{epsfig}
\usepackage{amsmath}
\usepackage{amssymb}
\usepackage{bm}
\usepackage{hyperref}
\usepackage{slashed}
\newcommand{\be} {\begin{equation}}
\newcommand{\ee} {\end{equation}}
\newcommand{\ba} {\begin{eqnarray}}
\newcommand{\ea} {\end{eqnarray}}

\newcommand{\no} {\nonumber}

\newcommand{\Br} {\mathcal B}
\newcommand{\cR} {\mathcal R}
\newcommand{\cF} {\mathcal F}

\newcommand{\dd}{\textrm{d}}
\newcommand{\bra}[1]{\left\langle{#1}\right\vert}
\newcommand{\ket}[1]{\left\vert{#1}\right\rangle}

\newcommand{\GeV}{\text{ GeV}}

\usepackage{color}
\definecolor{darkblue}{cmyk}{1,0.3,0,0.2}
\definecolor{violet}{cmyk}{0,1,0,0.2}
\hypersetup{colorlinks, bookmarksnumbered, citecolor=darkblue, linkcolor=darkblue, pdfstartview=FitH, urlcolor=darkblue, linktocpage}

\newcommand{\arXhref}[1]{\href{http://arxiv.org/abs/#1}{#1}}
\journalname{Eur. Phys. J. C}

\begin{document}

\title{On the Standard Model predictions for $R_K$ and $R_{K^*}$}

\author{ Marzia Bordone\thanksref{addr2},   Gino Isidori\thanksref{addr2},   Andrea Pattori\thanksref{addr2,addr3}
}

\institute{Physik-Institut, Universit\"at Z\"urich, CH-8057 Z\"urich, Switzerland 
\label{addr2}
\and
Dipartimento di Fisica e Astronomia "G. Galilei", Universit\`a di Padova, Via Marzolo 8,
I-35131 Padua, Italy \label{addr3}}

\maketitle

\begin{abstract}
We evaluate the impact of radiative corrections in the ratios
$\Gamma[B\to M \mu^+\mu^-]/\Gamma[B\to M e^+e^-]$ 
when the meson $M$ is a $K$ or a $K^*$. Employing the cuts on  $m^2_{\ell\ell}$ 
and the reconstructed $B$-meson mass presently applied by the LHCb 
Collaboration, such corrections do not exceed a few~$\%$. Moreover, their effect 
is well described (and corrected for)~by existing Montecarlo codes.  
Our analysis reinforces the interest of these observables as clean probe of  
physics beyond the Standard Model.
\end{abstract}

\section{Introduction}
\label{intro}

The Lepton Flavor Universality (LFU) ratios
\be
R_M [q^2_{\rm min},~q^2_{\rm max}]  = \frac{  \displaystyle{\int_{q^2_{\rm min}}^{q^2_{\rm max}}  \dd q^2  \frac{\dd\Gamma (B\to M \mu^+\mu^-) }{\dd q^2 }} }{   \displaystyle{ \int_{q^2_{\rm min}}^{q^2_{\rm max}}  \dd q^2 \frac{\dd\Gamma (B\to M e^+ e^-) }{\dd q^2 }  }}~,
\ee
where $q^2=m^2_{\ell\ell}$, are   very clean probes of physics beyond the Standard Model 
(SM): they  have small theoretical uncertainties and  are 
sensitive to possible new interactions 
that couple in a non-universal way to electrons and muons~\cite{Hiller:2003js}. 
A strong interest in $R_K$ has recently  been raised by the LHCb result~\cite{Aaij:2014ora}
\begin{equation}
R_K \left[1\GeV^2,~6\GeV^2\right] =0.745^{+0.090}_{-0.074}\pm 0.036~,
\label{RK}
\end{equation}
that differs from the na\"ive expectation
\be
R_{K^{(*)}}^{(\rm SM)} =1
\label{eq:naive}
\ee
by about $2.6\sigma$. The interest is further raised by
the combination of this anomaly with other $b\to s \ell^+\ell^-$ 
observables~\cite{Altmannshofer:2015sma,Descotes-Genon:2015uva},
and by the independent hints of violations of LFU observed 
$B\to D^{(*)}\tau\nu_\ell$ decays~\cite{Lees:2012xj,Huschle:2015rga,Aaij:2015yra}.

While perturbative and non-perturbative QCD contributions cancel in $R_{K^{(*)}}$
(beside trivial kinematical factors), this is not necessarily the case 
for QED corrections.  In particular, QED collinear singularities  
induce corrections  of order $(\alpha/\pi) \log^2(m_B/m_\ell)$
to $b\to s \ell^+\ell^-$ transtions~\cite{Huber:2015sra,Huber:2005ig}
that could easily imply $10\%$ effects in  $R_{K^{(*)}}$.  The purpose of this paper is to estimate  
these corrections and to precisely quantify up to which level a deviation 
of $R_K$ or $R_{K^*}$ from 1 can be considered a clean signal of physics beyond the SM.

\section{QED corrections in $R_M$}

A complete evaluation of QED corrections to $B\to M \ell^+\ell^-$ decay amplitudes 
is a non-trivial task, due to the interplay of perturbative and non-perturbative dynamics
(see e.g.~\cite{Gratrex:2015hna}).
However, the problem is drastically simplified if we are only interested 
in the LFU ratios $R_M$, especially in the low  dilepton invariant mass  region,
and if interested in possible deviations from Eq.~(\ref{eq:naive}) exceeding  $1\%$.
In this case the problem is reduced to evaluating
 $\log(m_\ell)$ enhanced terms, whose origin can be unambiguously 
 traced to soft and collinear photon emission. The latter 
represents a universal correction factor~\cite{Yennie:1961ad,Weinberg:1965nx}
that can be implemented,  by means of appropriate convolution functions,\footnote{For a discussion about the 
implementation of universal QED  corrections in a general EFT context see also 
Ref.~\cite{Isidori:2007zt}. }
irrespective of the specific short-distance structure of the amplitude.

\subsection{Universal radiation function}
\label{eq:radiation} 

Following the above observation, the treatment of soft and collinear photon emission 
in $B\to M \ell^+\ell^-$ closely resemble that applied to $h\to 2e 2\mu$ decays 
in Ref.~\cite{Bordone:2015nqa}.  The key observable we are interested in is the differential  lepton-pair invariant-mass distribution 
\be
\cF_{M}^{\ell}( q^2 )=\frac{\dd\Gamma (B\to M \ell^+\ell^-) }{\dd q^2 }~.
\ee
The complete structure of infrared (IR) divergences in the decay is channel dependent~\cite{Gratrex:2015hna};
however, 
the $\log(m_\ell)$ enhanced terms can be factorized and are independent from the spin of the meson $M$.

The leading QED corrections can be unambiguously identified working in the limit of massless leptons, 
retaining only the mass terms regulating collinear singularities. 
In this limit we define the radiator $\omega(x,x_\ell)$, that represents the probability density function 
that a dilepton system retains a fraction $\sqrt{x}$ of its original invariant mass after brems\-strah\-lung.
Namely we define $x=q^2/q_0^2$, where $q_0^2$ is the initial dilepton invariant mass squared (pre brems\-strahlung),
and we introduce the variable $x_\ell = 2 m_{\ell}^{2}/ q_0^2 $ that regulates collinear singularities.
 In order to match the IR-safe observable directly probed in experiments, 
the integration range of $x$ is determined by  the requirement 
that the recons\-tructed $B$-meson mass ($m_{B}^{\rm rec}$), 
from the measurement of leptons and hadron momenta, 
is above a minimum value.

In order to regulate IR divergences, we introduce an (unphysical) IR-regulator $x_*$ ($x_* \ll 1$), 
defined as the minimal detectable value of of $1-x$. The full radiator $\omega(x,x_\ell)$
is then decomposed as 
\be
	\omega(x,x_\ell) = \omega_1(x,x_\ell) \theta(1 -  x-x_*) + \omega_2(x,x_\ell,x_*) \delta(1-x)~,
\ee
where the explicit form of  $\omega_{1,2}$  in the limit $(1-x) \ll 1$ and $x_\ell, x_*  \ll 1$ is 
\ba
	\omega_1(x,x_\ell) & =&    \frac{\alpha}{\pi} \frac{1}{1-x} \left[ -2 +  (1+x^2)  \log \left( \frac{ 2x}{x_\ell} \right) \right]~, 
	\no\\
	\omega_2(x,x_\ell,x_*) & =& 1 -
		\frac{\alpha}{\pi} \left\{ 
		 \frac{5}{4}- \frac{\pi^2}{3} + 2\log(x_*)  \right.  \no\\
		 &&\left.  \qquad\quad	+\left[  \frac{3}{2} +2 \log(x_*) \right] \log\left(\frac{x_\ell}{2}\right)  \right\}~.
\ea
The first term, $\omega_1$, describes the real emission of a photon such that the lepton pair retains a fraction $\sqrt{x}$ of its invariant mass; the $\theta$-function implements the corresponding IR regulator.
The second term, $\omega_2$, 
describes the events in which the soft radiation is below the IR regulator, as well as the effect of virtual corrections.

We have determined the  structure of $\omega_1$ by means of an explicit  $O(\alpha)$ calculation of the real emission,  
while $\omega_2$ has been determined by the condition 
\be
\omega_2(x,x_\ell,x_*) = 1 - \int_{2 x_\ell}^{1-x^*} \dd x~\omega_1(x,x_\ell)    
\ee
that, by construction, ensure the independence of the full radiator from the IR regulator 
and the normalization condition
\be
 \int_{2 x_\ell}^1 \dd x~\omega(x,x_\ell) =1~.
\ee
The latter is valid up to finite (non log-enhanced) 
corrections of $O(\alpha/\pi)$ that define the accuracy of our approximation.

We can thus write the double differential distribution in terms of  
the invariant mass of the  dilepton system before bremsstrahlung  
  and  $x =q^2 / q^2_{0}$ as 
\begin{align} \label{eq: full diff}
\frac{\dd^2 \Gamma}{\dd q^2_0  \dd x }  =  \cF_M^{(0)}( q^2_0) \omega(x,x_\ell, x_*)~,
\end{align}
where $\cF_M^{(0)}(q^2_0)$ denotes the non-radiative spectrum.
Starting from Eq.~(\ref{eq: full diff}) we can extract the 
double differential spectrum after radiative corrections. To this purpose,
we first trade $x$ for $q^2$,   we then   integrate over all the possible 
values of $q^2_0$ determined by the cut on $m_{B}^{\rm rec}$, 
namely\footnote{In principle, from a pure kinematical point of view, the cut 
on $m_{B}^{\rm rec}$ allow $q_0^2$ values even exceeding the bound in Eq.~(\ref{eq:boundq0}); however, this 
occurs only for non-soft and non-collinear emissions that are beyond our approximations.} 
\be
  q_0^2 \leq q_{0,{\rm max}}^{2}(q^2,\delta)  =  \frac{q^2}{ \delta^2}
  \left[ 1+ (1-\delta^2) \frac{m_M^2}{ m_B^2 \delta^2 -q^2} \right] ~, \label{eq:boundq0}
\ee
where $\delta = m_{B}^{\rm rec}/ m_B <1$.
Proceeding this way we finally obtain: 
\ba
\cF_M^\ell( q^2 ) =  \int_{q^2}^{ q_{0,{\rm max}}^{2} } \frac{\dd q_0^2 }{q_0^2}   \cF_M^{(0)}( q^2_0) 
\omega\left(\frac{ q^2 }{q_{0}^2}, \frac{2 m_{\ell}^{2} }{q_0^2} \right),
 \label{eq:Ffinal}
\ea

We stress that the result in Eq.~(\ref{eq:Ffinal}) 
includes both real and virtual QED corrections. The latter have been indirectly determined by the normalization condition for 
$\omega(x,x_\ell)$,
that is the same condition applied in showering 
algorithms~\cite{Davidson:2010ew}, and that follows from the safe IR behavior of 
the photon-inclusive dilepton spectrum. 

Before concluding this section, we summarize below the size of neglected contributions
and the accuracy of this calculation. 
\begin{itemize}
\item{} As anticipated, we do not control $O(\alpha/\pi)$ virtual corrections 
that are regular in the limit ${m_\ell\to0}$. The latter are expected to be safely 
below the $1\%$ level. 
\item{} The calculation of the real emission has been done in the limit $m_{\ell}^{2} \ll q^2$ that is certainly an excellent 
approximation in the electron case, while it is less good in the muon case; however, also in this case the 
neglected contributions  are $O(\alpha/\pi)$  non log-enhanced terms.
\item{} In the case of a charged meson in the final state, we should consider also the radiation from the meson leg.
We have checked by means of an explicit calculation at $O(\alpha)$ (employing a generic  hadronic matrix element) 
that the latter do not interfere with the radiation of the lepton legs at the leading-log level once we integrate over the 
leptonic angles.\footnote{This happens because the leptonic current carries an overall neutral electric charge.}  
The radiation of the meson leg can thus be considered separately by means of an independent radiation function.
A quantification of its effect in the $B^+ \to K^+ \ell^+\ell^-$ case is discussed in sect.~\ref{sect:num}.
\item{} Independently of the charge of the meson, an additional contribution to the real radiation is
due to structure-de\-pen\-dent terms (i.e.~separately gauge-invariant amplitudes that vanish 
in the $E_\gamma \to 0$ limit). By construction, these amplitudes are free from soft singularities
but could have collinear singularities. However, these vanishes after a symmetry integration over the 
leptonic angles for the same argument discussed above.
\item{}  In order to quantify the impact of radiative corrections we need a theoretical input for the 
non-radiative spectrum $\cF_M^{(0)}( q^2_0)$, whose explicit expression for $B\to K $ and $B\to K^*$ transitions is 
discussed in sect.~\ref{sect:nonrad}.  From   Eq.~(\ref{eq:Ffinal}) it is clear that, 
as long as $\cF_M^\ell( q^2)/\cF_M^{(0)}(q^2)$ is a smooth function of $q^2$, the relative impact of radiative 
corrections in $R_M$ is insensitive to the dynamics responsible for the $B\to M \ell^+\ell^-$ decay.
\end{itemize}
 
\subsection{Parameterization of the non-radiative spectrum}
\label{sect:nonrad}

The choice of the radiative spectrum for the $B\to K^+\ell^+\ell^-$ decay is quite simple. 
In full generality we can write 
\be
\cF_K^{(0)}( q^2 ) \propto \lambda^{3/2}  (q^2) \left| f_+(q^2) \right|^2 \left[  |a_9 (q^2) |^2 +  
|a_{10} |^2 \right]~,
\ee
where $\lambda(s) = (m_B^4 + m_K^4 + s^2 - 2 m_K^2 m_B^2  - 2 s m_B^2 - 2 s m_K^2)/m_B^4$,
$f_+(q^2)$ is the $B\to K$ vector form factor
\be
   \bra{K(k)} \bar{s} \gamma_\mu b \ket{B(p)} = f_+(q^2) (p+k)^\mu+ O(q^\mu)
\ee
and $a_{9} (q_0^2)$ and  $a_{10}$ denote the 
effective Wilson coefficients of the vector and the axial-vector  components 
of the leptonic current~\cite{Bartsch:2009qp}. For our numerical analysis we use the 
parameterization of the form factor and the numerical values of the Wilson
coefficients from Ref.~\cite{Bartsch:2009qp}. 

In order to provide an effective  description  of the non-perturbative 
distortion of the spectrum induced by the charmonium resonances, we modify the 
 vector effective Wilson coefficient as follows 
\be
 a_9 (q^2) =   a^{\rm pert}_9 (q^2)  +  \kappa_{\psi} \frac{q^2}{ q^2 - m^2_{\psi} +i  m _{\psi}~
 \Gamma_{\psi} }
 \label{eq:Jpsi}
\ee
where  $\{m_\psi, \Gamma_{\psi}\}$ are the experimental mass and width  of the $J/\psi(1S)$ 
state, and the value of the (real) effective coupling $\kappa_{\psi}$ has been fixed in order  
to reproduce  $\Br(B \to K \psi)$  in the narrow width approximation.
This description is certainly approximate (see e.g.~the discussion in 
Ref.~\cite{Kruger:1996cv,Lyon:2014hpa}), but it provides a good estimate of the region where 
the $B\to K^+\ell^+\ell^-$ spectrum starts to vary rapidly with $q^2$, that is  relevant 
in order to define the region of validity of our approach. 

As far as the $B\to K^*\ell^+\ell^-$ is concerned, we proceed introducing the standard 
set of vector, axial, and tensor form factors 
\ba
&& \bra{K^*} \bar{s} \gamma_\mu b \ket{B} =
 \frac{2 V(q^2)}{m_B + m_V} \varepsilon_{\mu\rho\sigma\tau} \epsilon^{*\rho} p^\sigma k^\tau ,\\
 &&   \bra{K^* } \bar{s} \gamma_\mu \gamma_5 b \ket{B }
         = i\epsilon^{*\rho}
            \left[ 2 m_V A_0(q^2) \frac{q_\mu q_\rho}{q^2}  \right. \no\\
&&\left.            + (m_B + m_V) A_1(q^2)\left(g_{\mu\rho} - \frac{q_\mu q_\rho}{q^2}\right)\right. \nonumber \\
     && - \left.A_2(q^2) \frac{q_\rho}{m_B + m_V}\left((p+k)_\mu - \frac{ \Delta m^2}{q^2} q_\mu\right) \right] ,\\
 &&   \bra{K^* } \bar{s} i \sigma_{\mu\nu} q^\nu b\ket{B )}
   = -2T_1(q^2) \varepsilon_{\mu\rho\sigma\tau} \epsilon^{*\rho} p^\sigma k^\tau, \\
  &&  \bra{K^*} \bar{s} i \sigma_{\mu\nu} \gamma_5 q^\nu b\ket{B }
      = iT_2(q^2) \left[ \epsilon^*_\mu \Delta m^2 - (\epsilon^* \cdot q) (p+k)_\mu\right] \nonumber
        \\
         && + iT_3(q^2) \left(\epsilon^* \cdot q\right)\left(q_\mu - \frac{q^2}{ \Delta m^2 }(p+k)_\mu\right) ,
\ea
where $ \Delta m^2 = m_B^2 - m_{K^*}^2$, whose numerical values are taken 
from Ref.~\cite{EOS} (and based on the original works in Ref.~\cite{Beneke:2001at}).
With these we proceed evaluating the differential rate  as, for instance, in Ref.~\cite{Hiller:2003js}. 

\begin{figure}[t]
\includegraphics[width=.45\textwidth]{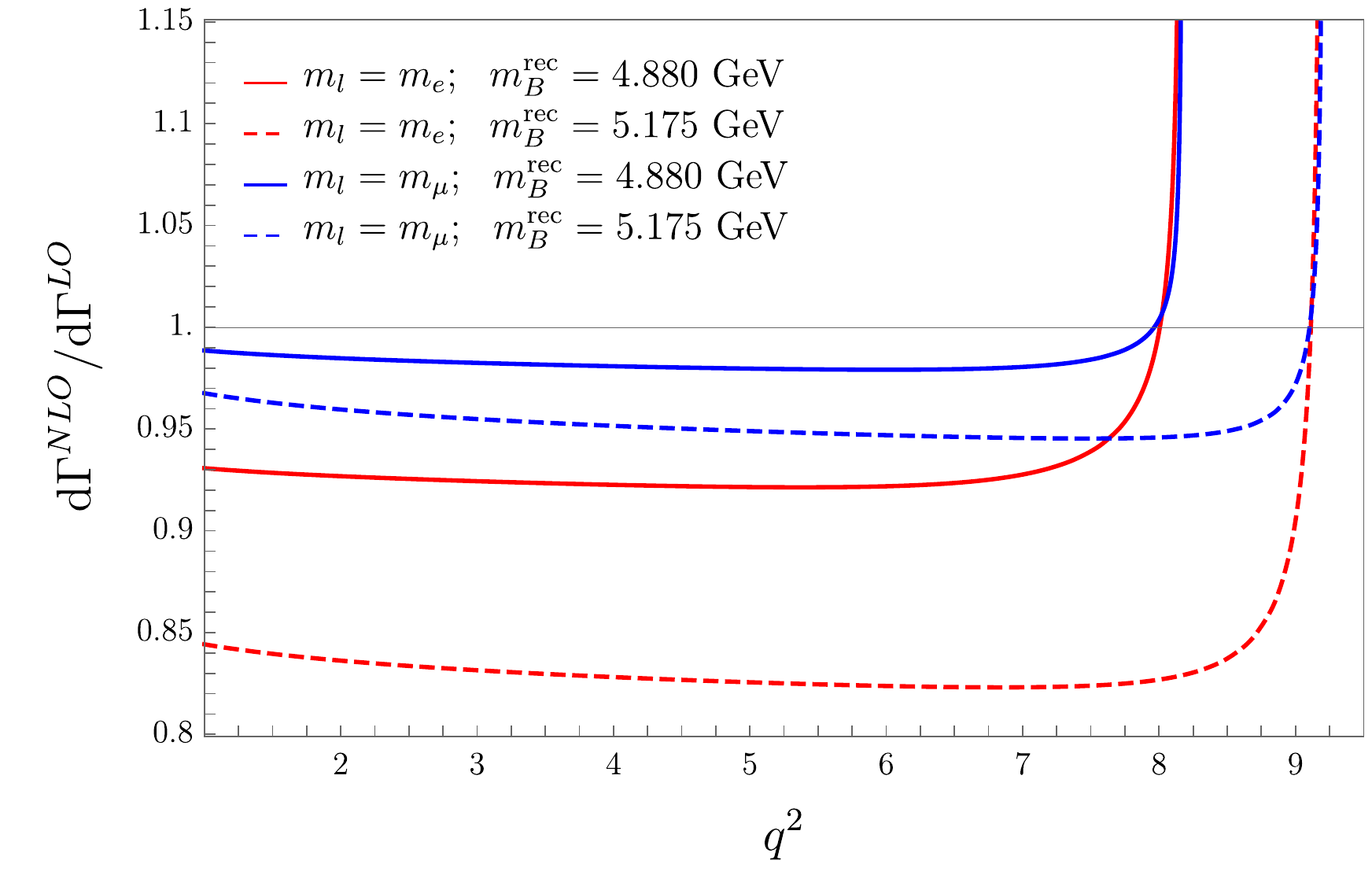}  
\caption{Relative impact of radiative correction in $B\to K^+\ell^+\ell^-$ decays 
for $q^2 \in [1,9.5]~{\rm GeV}^2$, with different 
cuts on the reconstructed mass and different lepton masses. 
\label{fig:one}}
\end{figure}

\section{Numerical results}
\label{sect:num}

The relative impact of radiative corrections in $B\to K^+\ell^+\ell^-$, namely a plot of the ratio
\be
\cR_K^\ell (q^2) = \frac{ \cF_K^\ell (q^2) }{ \cF_K^{(0)} (q^2)  },
\ee
 is shown in Fig.~\ref{fig:one} in the region $q^2 \in [1,9]~{\rm GeV}^2$.
The different colors correspond to different lepton masses 
(red for the electron  and blue for the muon).
Dashed and full lines correspond to different choices of the minimal cut on the 
reconstructed $B$-meson mass from the momenta of charged particles. We 
have choosen for the latter the two values used in Ref.~\cite{Aaij:2014ora} for the 
analysis of the electron modes ($m_{B}^{\rm rec} \geq 4.880$~GeV, full lines) 
or the  muon  modes ($m_{B}^{\rm rec} \geq 5.175$~GeV, dashed lines).

\begin{table}[t]
\centering
\begin{tabular}{|c|c|c|}
\hline
$B\to K \ell^+\ell^-$  & $ \ell= e$   & $\ell=\mu$   \\
\hline
$m_B^{\text{rec}}=4.880$ GeV & $-7.6\%$  &  $-1.8\%$ \\
\hline
$m_B^{\text{rec}}=5.175$ GeV &$-16.9\%$  & $-4.6\%$\\
\hline\hline 
$B\to K^*  \ell^+\ell^-$ &  $ \ell= e$   & $\ell=\mu$   \\
\hline
$m_B^{\text{rec}}=4.880$ GeV & $ -7.3\%$  &  $-1.7\%$ \\
\hline
$m_B^{\text{rec}}=5.175$ GeV &$- 16.7 \%$  & $-4.5\%$\\
\hline
\end{tabular}
\caption{Relative impact of radiative corrections  for $q^2\in [1,6]~{\rm GeV}^2$,
with different cuts on the reconstructed mass and different lepton masses.
\label{tab:one} }
\end{table}

\begin{figure}[t]
\includegraphics[width=.45\textwidth]{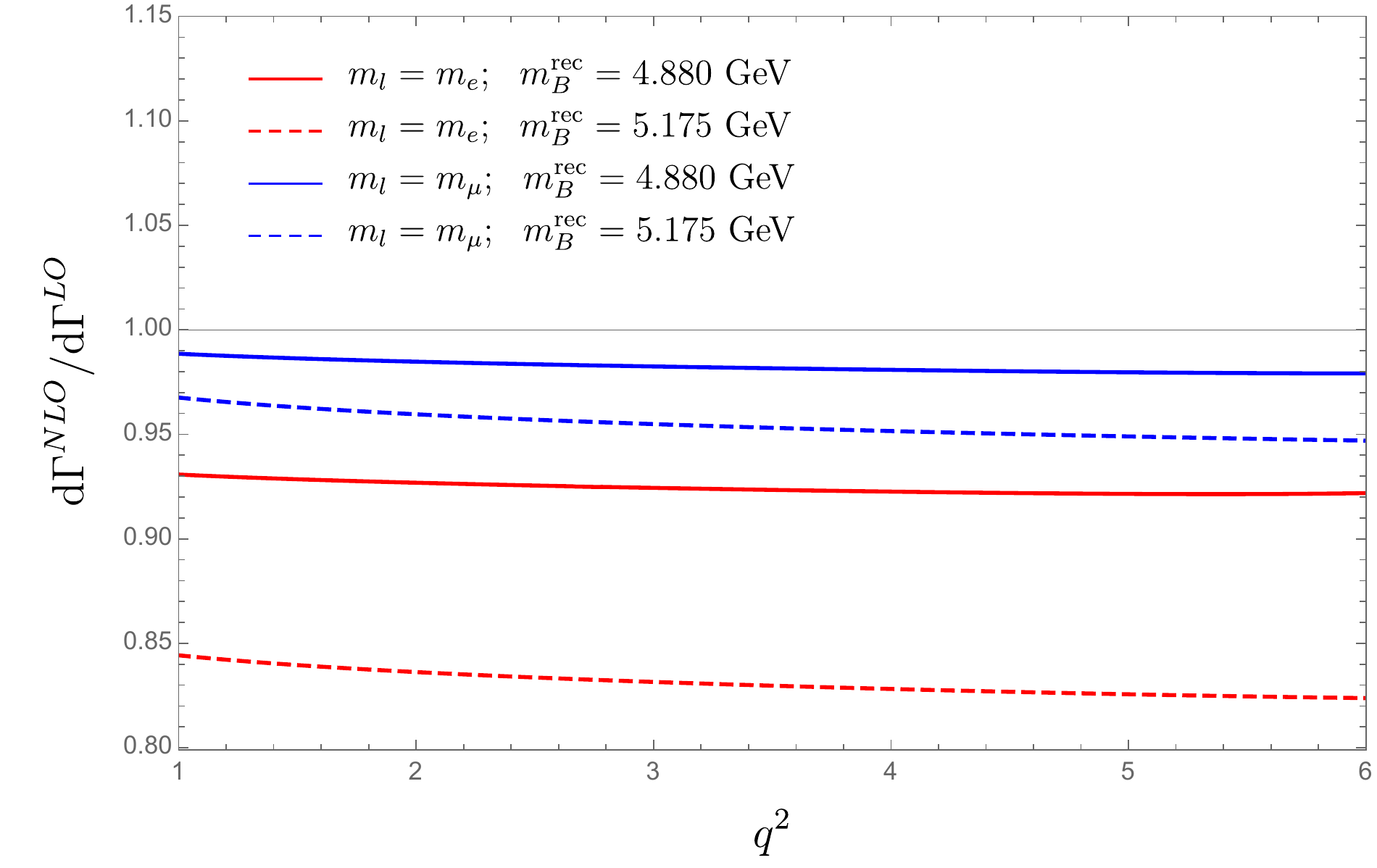} \\
\includegraphics[width=.45\textwidth]{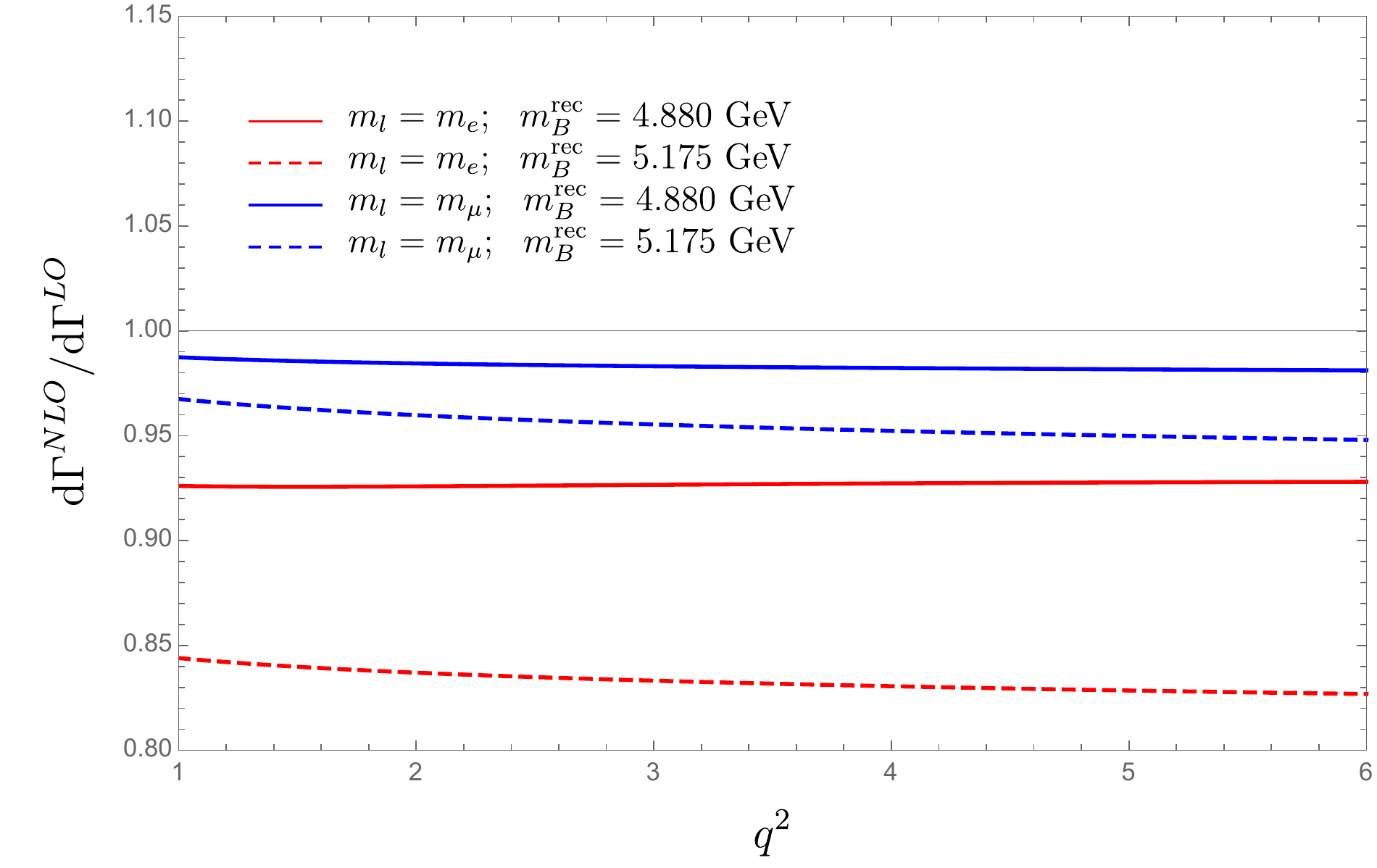}
\caption{Relative impact of radiative correction in $B\to K \ell^+\ell^-$ (up) and  
in $B\to K^* \ell^+\ell^-$  (down)  
for $q^2 \in [1,6]~{\rm GeV}^2$, with different 
cuts on the reconstructed mass and different lepton masses.
\label{fig:two}. }
\end{figure}

The first point to be noted in Fig.~\ref{fig:one} is that $\cR_K^\ell (q^2)$ is a 
smooth function for sufficiently low values of $q^2$, while a sudden rise appear 
close to the resonance region. The latter is a manifestation of the radiative return from
the $J/\Psi$ peak. The position where the $J/\Psi$ contamination appears depends 
only from the cut imposed on $m_{B}^{\rm rec}$. Even for the looser cut applied in the 
electron case the region $q^2 \in [1,6]~{\rm GeV}^2$ is free from the $J/\Psi$ contamination
and   can be estimated with good theoretical  accuracy (see Fig.~\ref{fig:two}).
To better quantify this statement we have explicitly checked that varying the phase of the 
effective coupling $\kappa_\psi$ in Eq.~(\ref{eq:Jpsi}) leads to per-mill modifications to $\cR_K^\ell (q^2)$
for $q^2 \leq 6$~GeV$^2$. We also have explicitly checked that the cut on $m_{B}^{\rm rec}$
eliminates photons from the  $J/\Psi$ peak
 also when considering the full kinematics of the event, 
i.e.~beyond the soft and collinear approximation on which we derived Eq.~(\ref{eq:boundq0}).

The second point to be noted is that in the regular region of the spectrum radiative 
corrections reach (or even exceed) the $10\%$ level for the electrons (as naively expected);
however, the net effect in $R_K$ is  significantly smaller. Indeed  the 
magnitude of the corrections is larger for electron vs.~muons, but it increases 
for $m_{B}^{\rm rec} \to m_B$. This imply that the specific choice of $m_{B}^{\rm rec}$ cuts 
applied by the LHCb collaboration, i.e.~a loose cut for the electrons and a tighter 
cut for the muons, give rise to a natural compensation of the QED corrections to $R_K$.  
  
The integrated corrections that quantity the modifications to $R_K$ are reported in Table~\ref{tab:one}.
Given the choice of $m_{B}^{\rm rec}$ applied in Ref.~\cite{Aaij:2014ora}, we estimate 
that radiative corrections induce a {\em positive} shift of the central value of $R_K$
of a about $\Delta R_K = +3\%$. This effect is taken into account by the LHCb collaboration,
who estimated the impact of radiative corrections with PHOTOS~\cite{Davidson:2010ew}, and properly corrected 
for in the result reported. We have explicitly checked that our estimate of  $\Delta R_K$ 
is in agreement with that obtained with PHOTOS up to differences within $\pm 1\%$.\footnote{We thank 
Rafael Silva Coutinho for a detailed comparison about the radiative corrections implemented in the LHCb 
analysis of $R_K$.}
 
In order to check the smallness of the non-$\log(m_\ell)$ enhanced terms,
in Table~\ref{tab:two} we report the effect of the radiation from the meson leg, that is 
IR divergent but has no collinear singularities. We evaluated these terms 
developing the corresponding radiator function (see Ref.~\cite{Isidori:2007zt}),
whose implementation depend only on $m_{B}^{\rm rec}$. As can be seen 
from Table~\ref{tab:two}, the results are well below the $1\%$ level.

\begin{table}[t]
\centering
\begin{tabular}{|c|c|}
\hline
$m_B^{\text{rec}}=4.880$ GeV & $-0.02\%$ \\
\hline
$m_B^{\text{rec}}=5.175$ GeV & $-0.18\%$  \\
\hline
\end{tabular}
\caption{Relative contribution of radiative corrections due emission 
from the meson leg, in the $B^+ \to K^+  \ell^+\ell^-$ case, for 
$q^2 \in [1,6]~{\rm GeV}^2$.
\label{tab:two}}
\end{table}

The impact of radiative corrections in the $B\to K^* \ell^+\ell^-$ decays is shown 
in Fig.~\ref{fig:two} and summarized by the integrated values reported  in Table~1.
The situation is very similar to the  $B^+ \to K^+  \ell^+\ell^-$ :
employing the same $m_{B}^{\rm rec}$  cuts for electron and muon 
modes as in Ref.~\cite{Aaij:2014ora}, we find that the net impact of 
radiative corrections is $\Delta R_{K^*} = +2.8\%$. Also in this case  
this effect is well described by PHOTOS and therefore can be 
properly corrected for  in future experimental  analyses.

\section{Conclusions}

The experimental result in  Eq.~(\ref{RK})
has stimulated a lot of theoretical activity~\cite{ Alonso:2014csa, Sahoo:2015pzk, Carmona:2015ena,
Hiller:2014yaa,Ghosh:2014awa,Biswas:2014gga,Hurth:2014vma,
Glashow:2014iga,Crivellin:2015mga,Sierra:2015fma,Crivellin:2015era,Celis:2015ara, Gripaios:2014tna,Becirevic:2015asa,Varzielas:2015iva,Alonso:2015sja,Calibbi:2015kma, Barbieri:2015yvd,  Crivellin:2015hha,   
Bauer:2015knc,Fajfer:2015ycq,Greljo:2015mma,
Guadagnoli:2015nra,Sahoo:2015qha,Falkowski:2015zwa,Pas:2015hca,Bhattacharya:2014wla,
Buttazzo:2016kid, Capdevila:2016ivx}
In view of this result and, especially, in view of possible  future experimental improvements
in the determination of  $R_K$ or $R_{K^*}$, we have re-examined the SM predictions of these LFU ratios. 

As we have shown,   $\log(m_\ell)$-enhanced 
QED corrections may induce sizable deviations from Eq.~(\ref{eq:naive}),
even up to $10\%$, 
depending on the specific cuts applied to define physical observables. 
In particular, a key role is played by the cuts on $q^2=m^2_{\ell\ell}$ and 
on the reconstructed $B$-meson mass. The former is important to avoid 
rapidly varying regions in the dilepton spectrum (where the theoretical tools 
to compute QED corrections become unreliable), while the latter defines the 
physical IR cut-off of the rates.  
Employing the cuts  pre\-sently applied by the LHCb 
Collaboration,  the corrections in $R_K$ do not exceed $3\%$. Moreover, their effect 
is well described (and corrected for in the experimental analysis) 
by existing Montecarlo codes.

According to our analysis, a deviation of $R_K$ or 
$R_{K^*}$ from 1 exceeding the $1\%$ level, performed along the lines of 
Ref.~\cite{Aaij:2014ora}  in the region  
 $1\GeV^2 < q^2 < 6\GeV^2 $,
would be a clear signal of physics beyond the Standard Model.

\begin{acknowledgements}
We thank Rafael Silva Coutinho and Nicola Serra for  useful discussions
about the LHCb analysis of $R_K$, and  Danny van Dyk for help with the 
numerical implementation of $B\to K^*$ form factors.  
This research was supported in part by the Swiss National Science Foundation 
(SNF) under contract 200021-159720.
\end{acknowledgements}

\bigskip

\noindent 
{\bf Added note}:  {\em SM predictions for $R_{K^*}$ in the low $q^2$ region.} \\

\noindent 
After this paper was published, the LHCb Collaboration has announced a measurement of $R_{K^*}$
in two bins in the  low $q^2$ region, reporting also in this case a significant deviation from 
unity~\cite{LHCbtalk}:\footnote{~The two values between square brackets in $R_{K^\ast}$
denote the $q^2$ range in GeV$^2$.}
\begin{align}
R_{K^\ast} [  0.045,~1.1]  &= 0.660^{+0.110}_{-0.070}\pm0.024~, \quad
 \nonumber \\[0.2cm]
R_{K^\ast} [  1.1,~6.0] & = 0.685^{+0.113}_{-0.069}\pm0.047~.  \quad 
\end{align}
Given the interest of these results, we provide here precise SM predictions for 
$R_{K^*}$ in these two bins, taking into account QED radiative corrections. 

The prediction in the  $1.1~\text{GeV}^2 \leq q^2  \leq 6~\text{GeV}^2$ bin 
does not differ from what discussed above. For the sake of clarity, 
we predict 
\begin{equation}
R_{K^\ast} [  1.1,~6.0]^{\rm SM}  =  1.00 \pm 0.01_{\rm QED}~,
\end{equation}
together with
\begin{equation}
R_{K^+} [  1.0,~6.0]^{\rm SM}  =  1.00 \pm 0.01_{\rm QED}~,
\end{equation}
for the extrapolated photon-inclusive observables reported by LHCb.
The subscript on the errors signals that the origin of this theoretical uncertainty are QED effects.
In this region the residual uncertainty due to form-factor errors (in absence of radiative corrections)
is negligible for both $K$ and $K^*$ modes. 

The prediction in the $0.045~\text{GeV}^2 \leq q^2  \leq 1.1~\text{GeV}^2$ bin is more delicate. 
The kinematical threshold of the muon mode, and the rapid (and flavour non-universal) variation of 
$d\Gamma/dq^2$ close to this threshold, imply larger theoretical uncertainties. 
First of all, even in absence of QED corrections, form-factor 
uncertainties do not cancel completely in this region. We estimate the latter to 
induce a $\pm 0.02$ error (in agreement with Ref.~\cite{Capdevila:2017bsm}).

\begin{figure}[t]
\includegraphics[width=.45\textwidth]{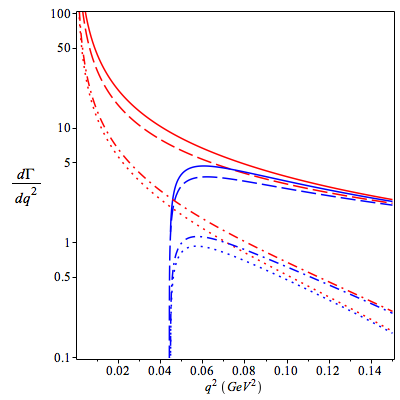}  
\caption{Contributions to $d\Gamma[B\to K^* \ell^+\ell^-(\gamma)]/dq^2$
(in arbitrary units) in the low $q^2$ region for $\ell=e$ (red) 
and $\ell=\mu$ (blue), before any cut in $m_B^{\text{rec}}$.
The full line is the photon-inclusive rate; 
the dashed line is the non-radiative FCNC rate; dotted and dash-dotted lines denote 
the contribution to the photon inclusive rate from $B\to K^* + \eta (\to \ell^+\ell^- \gamma)$
with (dash-dotted) and without (dots) interference with the soft radiation from the FCNC frate. 
\label{fig:lowq2}}
\end{figure}

As far as QED corrections are concerned, a specific aspect of the near-threshold region is the sensitivity 
to light-hadron effects. Non-negligible extra contributions to the pho\-ton-inclusive rate are obtained
by direct-emission amplitudes of the type $B \to K^* P^0 \to K^* \ell^+\ell^-\gamma$, 
where $P^0$ denotes an almost on-shell  $\eta$ or $\pi^0$ state. The $\eta$-mediated contribution 
turns out to be particularly sizeable given $\Br(B \to K^* \eta) \approx 1.6 \times 10^{-5}$ and 
$\Br(\eta \to e^+e^-\gamma) \approx 0.7 \%$.\footnote{
	In absence of a lower cut on $q^2$ and $m_B^{\text{rec}}$, the rate for 
$B \to K^* \eta \to K^* e^+ e^-\gamma$ is 
about $30\%$ of  $\Gamma(B \to K^* e^+ e^-;~q^2 < 0.1)$.}
An illustration of the impact of the latter is shown in 
Fig.~\ref{fig:lowq2}. 

Some comments on the light-hadron contribution are in order:
\begin{itemize}
\item[i.] This contribution is an irreducible part of 
the photon-inclusive rate (which is the only well-defined physical observable) 
and, as such, it must be included in the theoretical prediction of $R_{K^*}$
(in the relevant kinematical region).
\item[ii.]  The leading effect 
is necessarily a {\em decrease} of $R_{K^*}$ compared to the non-radiative case
(the radiative tails of electron and muon modes are both enhanced, but the effect is 
smaller in the muon case given the proximity to the phase-space border).
The decrease of $R_{K^*}$ is further enhanced by the looser $m_B^{\text{rec}}$ 
cut on electron  vs.~muon modes.
\item[iii.] There is a non-negligible interference between the meson-mediated 
amplitude and the soft-photon emission of the genuine FCNC  amplitude. This interference  
induces a (theoretical) uncertainty  in estimating this effect given the unknown relative phases of the amplitudes.
An additional source of uncertainty is provided by any other contribution of the type 
$B \to K^* \gamma+ \gamma^*(\to e^+e^-)$, for which we do not have 
a reliable normalization.
\item[iv.]  Above the threshold region also 
the meson-mediated amplitude becomes lepton universal (Fig.~\ref{fig:lowq2}),
and the uncertainty of this contribution becomes negligible for  $q^2 > 0.1~\text{GeV}^2$.
\end{itemize}
Taking into account the kinematical cuts $m_B^{\text{rec}}=4.500$ GeV (for $\ell=e$) and
$m_B^{\text{rec}}=5.150$ GeV (for $\ell=\mu$), we estimate the meson-mediated 
contribution to yield\footnote{The result in Eq.~(\ref{eq:shift})
holds under the assumption that any contribution to the photon-inclusive 
electron rate with $q^2 < 0.045~\text{GeV}^2$ is subtracted 
(or corrected for) 
on the experimental side, otherwise the correction could be significantly larger.}
\be
\Delta_{\rm QED} R_{K^*} [0.045,~ 1.1]\approx -0.017~.
\label{eq:shift}
\ee 
Given the discussion above, 
we assign a conservative $\pm 0.02$ error to the whole QED corrections in this region.
Our final SM estimate is then
\begin{eqnarray}
R_{K^\ast} [  0.045 , 1.1 ]^{\rm SM}  &=&  0.906 \pm 0.020_{\rm QED}  \pm 0.020_{\rm FF} \nonumber \\
&=& 0.906 \pm 0.028_{\rm th}~.
\label{eq:verylowq2}
\end{eqnarray}
It must be stressed that the (relatively) large theoretical uncertainty in  
(\ref{eq:verylowq2}) is due to the definition of the bin, that starts at the 
di-muon threshold. Setting the lower threshold to 0.1~GeV$^2$ (a value that we advocate in 
view of future experimental analyses) we find 
\bigskip
\begin{eqnarray}
R_{K^\ast} [  0.1,   1.1 ]^{\rm SM}  &=&  0.983 \pm 0.010_{\rm QED}  \pm 0.010_{\rm FF} \nonumber \\
&=& 0.983 \pm 0.014_{\rm th}~.
\end{eqnarray}

\end{document}